\documentclass{article}
\usepackage{arxiv}
\usepackage[utf8]{inputenc} 
\usepackage[T1]{fontenc}    
\usepackage{hyperref}       
\usepackage{url}            
\usepackage{booktabs}       
\usepackage{amsfonts}       
\usepackage{nicefrac}       
\usepackage{microtype}      
\usepackage{lipsum}		
\usepackage{graphicx}
\usepackage{doi}

\usepackage{cite}
\usepackage{amsmath,amssymb,amsfonts}
\usepackage{algorithmic}
\usepackage{algorithm}
\usepackage{multirow}
\usepackage{subcaption}
\usepackage{graphicx}
\usepackage{textcomp}
\usepackage{hyperref}
\usepackage{xcolor}
\usepackage{tikz}
\usetikzlibrary{matrix, positioning,backgrounds}

\usepackage{pgfplots}
\pgfplotsset{compat=1.17}

\title{AndroIDS : Android-based Intrusion Detection System using Federated Learning}

\author{ \href{https://orcid.org/0000-0002-7734-0367}{\includegraphics[scale=0.06]{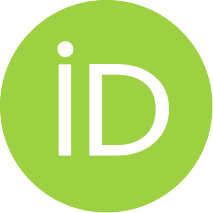}\hspace{1mm}Akarsh K Nair}\thanks{This work is partially supported by the US National Science Foundation grant 2431531.} \\
	Department of Computer Science Engineering\\
	IIIT Kottayam\\
	Kottayam, India \\
	\texttt{akarshkn@iiitkottayam.ac.in} \\
	\And
	\href{https://orcid.org/0000-0000-0000-0000}{\includegraphics[scale=0.06]{orcid.pdf}\hspace{1mm}Shanik Hubert Satheesh Kumar} \\
	Department of Computer Science Engineering\\
	IIIT Kottayam\\
	Kottayam, India \\
	\texttt{shanikhubert@gmail.com} \\
    \And
	\href{https://orcid.org/0000-0000-0000-0000}{\includegraphics[scale=0.06]{orcid.pdf}\hspace{1mm}Deepti~Gupta} \\
	Department of Computer Information Systems\\
	Texas A\&M University-Central Texas\\
    Texas, USA\\
	 \texttt{d.gupta@tamuct.edu}}

\renewcommand{\shorttitle}{AndroIDS}

\hypersetup{
pdftitle={A template for the arxiv style},
pdfsubject={q-bio.NC, q-bio.QM},
pdfauthor={David S.~Hippocampus, Elias D.~Striatum},
pdfkeywords={Federated learning, IDS, mobile IoT , privacy
and security},
}

\begin{document}
\maketitle

\begin{abstract}
	The exponential growth of android-based mobile IoT systems has significantly increased the susceptibility of devices to cyberattacks, particularly in smart homes, UAVs, and other connected mobile environments. This article presents a federated learning-based intrusion detection framework called AndroIDS that leverages system call traces as a personalized and privacy-preserving data source. Unlike conventional centralized approaches, the proposed method enables collaborative anomaly detection without sharing raw data, thus preserving user privacy across distributed nodes. A generalized system call dataset was generated to reflect realistic android system behavior and serves as the foundation for experimentation. Extensive evaluation demonstrates the effectiveness of the FL model under both IID and non-IID conditions, achieving an accuracy of 96.46\% and 92.87\%, and F1-scores of 89\% and 86\%, respectively. These results highlight the model’s robustness to data heterogeneity, with only a minor performance drop in the non-IID case. Further, a detailed comparison with centralized deep learning further illustrates trade-offs in detection performance and deployment feasibility. Overall, the results validate the practical applicability of the proposed approach for secure and scalable intrusion detection in real-world mobile IoT scenarios.
\end{abstract}


\section{Introduction}
The extensive growth of mobile Internet of Things (IoT) devices, ranging from smartphones and tablets to smart TVs and wearable systems, has transformed the digital environment across various industries such as smart cities, industrial systems, and healthcare infrastructure. A significant portion of these devices utilize the android operating system due to its open-source nature and adaptable functionality. However, such extensive deployment of android has increased its exposure to adversarial threats, making attack and intrusion detection a crucial component for securing mobile IoT systems.

Traditional Intrusion Detection Systems (IDS) rely heavily on centralized data collection and processing, which poses significant privacy, scalability, and communication overhead challenges, especially in light weight distributed IoT networks. Furthermore, as these devices generate vast amounts of system-level data, such as in the form of system calls, centralized collection not only puts on load network infrastructure but also raises concerns over data confidentiality and user privacy. Recent advancements in Federated Learning (FL) offer a promising solution to these limitations. By enabling collaborative model training across devices without exposing raw data, FL supports privacy-aware model development in distributed settings. When combined with system call analysis, the model can capture intricate behavioral patterns of applications and system-level activities, thus facilitating an efficient FL-based IDS for detecting anomalies and intrusions while preserving user privacy.

This article explores the formulation and analysis of a lightweight intrusion detection framework based on FL using android system call traces. The custom developed dataset represents a generalized perspective of android-based device behavior, thus making it broadly applicable to a wide range of mobile IoT systems. The performance of the proposed \textit{AndroIDS} framework is compared with a traditional centralized machine learning model, providing empirical insights into standard evaluation parameters such as detection accuracy and deployment feasibility in real-world settings.

The contributions of the article are as follows:
\begin{itemize}
\item An FL-based intrusion detection framework leveraging android system calls for identifying malicious activity across distributed mobile IoT devices is formulated.
\item A generalized dataset comprising of system call sequences collected from android environments, reflecting realistic mobile device behavior is presented.
\item A detailed comparative analysis between centralized and federated training setup in IID and non IID is conducted, evaluating detection efficacy and robustness. 
\item The practical applicability of the \textit{AndroIDS} framework in mobile IoT use cases is discussed with a special emphasis on smart home devices, industrial IoT platforms, and UAV systems operating on android-based environments.
\end{itemize}

The rest of this paper is organized as follows: Section~\ref{fourmuatiol} presents the problem formulation followed by section~\ref{releated} presenting related work. Section~\ref{systemmodel} details the proposed methodology and system design. Section~\ref{results} provides experimental results and  Section~\ref{discussions}  presents the discussions and futurescope. Section~\ref{concluion}
concludes the paper with future directions.
\section{Problem Formulation}\label{fourmuatiol}
The goal is to collaboratively generate global binary classifier $f(.)$ over a set of distributed clients without centralizing raw data. Each client $k$ locally trains the model, minimising a loss function $\mathcal{L}_k$ based on its private dataset $\mathcal{D}_k$. The central server aggregates local model updates via classical averaging using the Federated Averaging  algorithm.

\begin{equation}
\min_{w} \sum_{k=1}^{K} \frac{n_k}{n} \mathcal{L}_k(w)
\label{eq:global_opt}
\end{equation}
where $w$ are the model parameters,
 ${n_k}$ is the number of local samples, and $\mathcal{L}_k(w)$ is the local empirical loss at client $k$.

Each client optimizes its local model over using local system call sequences processed using TF-IDF and trained using a GRU-based architecture. The local objective is defined as:
\begin{equation}
\mathcal{L}_k(w) = \frac{1}{n_k} \sum_{i=1}^{n_k} \ell(f(x_i; w), y_i)
\label{eq:local_loss}
\end{equation}

The model updates are then aggregated by the server using the FedAvg~\cite{pmlr-v54-mcmahan17a} algorithm:
\begin{equation}
w_{t+1} = \sum_{k=1}^{K} \frac{n_k}{n} w_t^k
\label{eq:fedavg}
\end{equation}

\section{Related Work}\label{releated}
Intrusion detection using system call analysis has been extensively studied as an effective approach for capturing behavioral anomalies in operating systems. This section reviews existing literature, focusing on critical areas: IDA using system calls, FL based IDS and light weight FL models.

\subsection{Intrusion Detection Using System Calls}
System call analysis remains a core technique for behavioral intrusion detection, as it captures fine-grained system-level interactions of programs. One such article presented in~\cite{9314111} proposed an IDS that combines system call sequences, text classification, and graph-based analyses to model the global behavior of applications. A modified system call graph is used to unify multiple detection techniques, and a deep neural network aggregates their outputs. Results across three datasets demonstrated improved detection rates and reduced false positives.

Zhang et al.~\cite{10.1145/3538969.3544413} proposed a similar framework for early malware detection in android apps. The proposed model uses TF-IDF and six ML classifiers, achieved an average accuracy of 99.34\% using 3000 system calls and was validated via a real-time client-server android deployment. Another HIDS is presented in~\cite{10.1145/3461462}, which introduces a deep learning-based anomaly detection method using WaveNet and recurrent neural networks. Predictions are aggregated at the application level for anomaly detection, and results on the PLAID dataset show significant improvements over prior techniques.

Despite these advancements, most existing techniques focus on either static setups~\cite{9191626} or narrowly defined datasets. Our work addresses this gap by using a more comprehensive android system call dataset, aiming for an IDS applicable across a range of mobile IoT devices and attack scenarios.

\subsection{Federated Learning for Intrusion Detection}

FL has emerged as a promising privacy-preserving approach for collaborative intrusion detection. One such approach is the FELIDS framework~\cite{10032055}, targetting agricultural-IoT infrastructures. FELIDS leverages local model training with deep learning models to detect attacks while maintaining data privacy. Evaluations on CSE-CIC-IDS2018, MQTTset, and InSDN demonstrated that FELIDS outperforms centralized approaches in both accuracy and privacy. To further optimize performance in federated settings, Friha et al.~\cite{FRIHA202217} proposed DAFL, a dynamic weighted aggregation framework. DAFL introduces filtering and weighting strategies for local models to improve detection performance while reducing communication overhead. It also offers strong scalability and low communication cost, making it suitable for bandwidth-constrained environments. Further extending FL-based IDS, Idrissi et al.~\cite{IDRISSI2023121000} proposed Fed-ANIDS, a federated anomaly-based IDS that computes intrusion scores using reconstruction errors from standard, variational, and adversarial autoencoders. Evaluated on multiple benchmark datasets, Fed-ANIDS demonstrated high detection accuracy and low false alarm rates.

Eventhough these frameworks showcase the benefits of FL for intrusion detection, they primarily focus on network traffic datasets (e.g., flow records or IoT sensor data) and often overlook host-level behavioral data such as system calls. Moreover, few studies specifically focus on FL in android-based environments, where device heterogeneity (ranging from OS versions to app behavior) poses unique challenges. Our work addresses this gap by applying FL to an android system call dataset, integrating the privacy-preserving benefits of FL with the high level behavioral analysis of system call-based intrusion detection.
\subsection{ Lightweight FL Models for Edge and Mobile IoT Devices}
A practical FL-based IDS for mobile and IoT devices must be computationally lightweight and adaptable to resource-constrained environments. Recent research has therefore focused on optimizing FL algorithms for edge deployment, with an emphasis on efficiency, privacy, and robustness. One such framework, LSFL was proposed in~\cite{9947081} as a lightweight and secure mechanism tailored for edge computing. LSFL introduces a two-server secure aggregation protocol that ensures both data privacy and Byzantine robustness, effectively preventing malicious nodes from influencing the global model.

In~\cite{9614340}, a lightweight privacy-preserving FL scheme was presented, specifically targeting resource constrained IoT devices. The framework uses parameter masking with a secret-sharing mechanism to secure local updates and introduces a secure mask reusing strategy to minimize communication overhead across training rounds. Article ~\cite{9354925} further investigates resource-efficient FL in mobile edge computing by categorizing optimization strategies into black-box and white-box approaches. A neural-structure-aware resource management framework is proposed, where mobile clients are assigned different subnetworks of a global model based on their local resource availability.

In the context of malware detection, FED-MAL~\cite{9984824} presents a federated framework for IoT devices that converts malware binaries into images and uses a compact CNN model, AM-NET, for classification. To improve generalizability under non-IID conditions, the method incorporates adversarial training at the edge across distributed clients. However, many of these studies assume fixed device configurations or operate in controlled environments. In contrast, our work aims to present a more generalizable FL-driven IDS for android. The study utilizes a broad, representative system call dataset and compare centralized and federated training to evaluate trade-offs in performance and efficiency.

\section{System Model}\label{systemmodel}

This section outlines the centralized deep learning baseline and the FL setup used to detect intrusions from android system call sequences in mobile IoT environments.

\subsection{Centralized Deep Learning Baseline}
\subsubsection{Dataset and Preprocessing}

The dataset comprises system call logs collected from benign and malicious android applications. Each log contains time-stamped system-level calls. A total of 5832 labeled files (2474 benign, 3358 malicious) were used. Pre-processing included: Removal of timestamps using regular expressions and token sequence formation by joining syscalls into space-separated strings. Further, TF-IDF vectorization was performed using a vocabulary of 1000 most frequent terms. In addition to model sequential behavior, a sliding window approach was employed, maintaining a window size of 10 and a stride of 2. The combination of sliding window and striding transforms the dataset into overlapping sequences suitable for temporal modeling. To enhance robustness and prevent overfitting, a standard noise injection mechanism using Gaussian noise (with standard deviation=0.02 ) was injected into the input vectors during training. The standard 70:30 ratio is maintained for training and validation split during training.

\subsubsection{Model Architecture}

The model was built using a lightweight GRU-based architecture. It consists of two GRU layers with 32 and 16 units respectively, followed by batch normalization and a dropout layer with a dropout rate of 0.5. Towards the final classification layers, the model uses two layers with the last layer employing sigmoid activation for final classification. The model employed the AdamW optimizer with a learning rate of 5$e^{-5}$ and trained with binary cross-entropy loss. Further, an early stopping mechanism has been initialized, triggering a halt when both training and validation accuracies exceeded 90\%.

\subsection{Federated Learning Setup}
\subsubsection{Client setup and Data Partitioning}
The FL system was implemented using the Flower framework, simulating three different combinations with three, six, and ten clients. For data distribution, IID and non-IID patten were followed. In the IID setting, the training dataset was randomly and evenly partitioned across all clients without class imbalance, ensuring statistical similarity across local datasets. In non-IID setting, label skew and sample skew were introduced, such that client side data distribution resembled real-world mobile IoT deployments, where different devices may experience distinct operational behaviors or threat profiles. 
\begin{figure}[ht]
  \centering
   \begin{tikzpicture}
    \begin{axis}[
        ybar stacked,
        bar width=10pt,
        width=0.9\linewidth,
        height=5.75cm,
        enlarge x limits=0.1,
        ymin=0,
        ymax=150,
        ylabel={Total Samples},
        xlabel={Client ID},
        symbolic x coords={C1,C2,C3,C4,C5,C6,C7,C8,C9,C10},
        xtick=data,
        nodes near coords,
        nodes near coords align={vertical},
        every node near coord/.append style={font=\footnotesize\bfseries, black},
        title={IID Distribution}
    ]
    \addplot+[fill=blue!30] coordinates {(C1,65) (C2,62) (C3,66) (C4,60) (C5,61) (C6,63) (C7,59) (C8,64) (C9,67) (C10,62)};
    \addplot+[fill=red!40] coordinates {(C1,60) (C2,63) (C3,61) (C4,65) (C5,64) (C6,62) (C7,66) (C8,61) (C9,60) (C10,64)};
    \end{axis}
    \end{tikzpicture}

    \vspace{0.15cm}

    \begin{tikzpicture}
    \begin{axis}[
        ybar stacked,
        bar width=10pt,
        width=0.9\linewidth,
        height=5.75cm,
        enlarge x limits=0.1,
        ymin=0,
        ymax=485,
        ylabel={Total Samples},
        xlabel={Client ID},
        symbolic x coords={C1,C2,C3,C4,C5,C6,C7,C8,C9,C10},
        xtick=data,
        nodes near coords,
        nodes near coords align={vertical},
        every node near coord/.append style={font=\footnotesize\bfseries, black},
        legend style={at={(0.02,0.98)}, anchor=north west, draw=none, fill=none},
        title={non-IID Distribution}
    ]
     \addplot+[fill=blue!30] coordinates {
      (C1,40) (C2,30) (C3,122) (C4,61) (C5,45)
      (C6,91) (C7,153) (C8,122) (C9,34) (C10,301)
    };
    \addplot+[fill=red!40] coordinates {
      (C1,61) (C2,71) (C3,30) (C4,40) (C5,15)
      (C6,112) (C7,152) (C8,81) (C9,36) (C10,158)
    };
    \legend{Benign, Malicious}
    \end{axis}
    \end{tikzpicture}
\caption{Client-wise sample and class distribution under IID and non-IID partitioning for FL training }
  \label{fig:data_dist}
\end{figure}
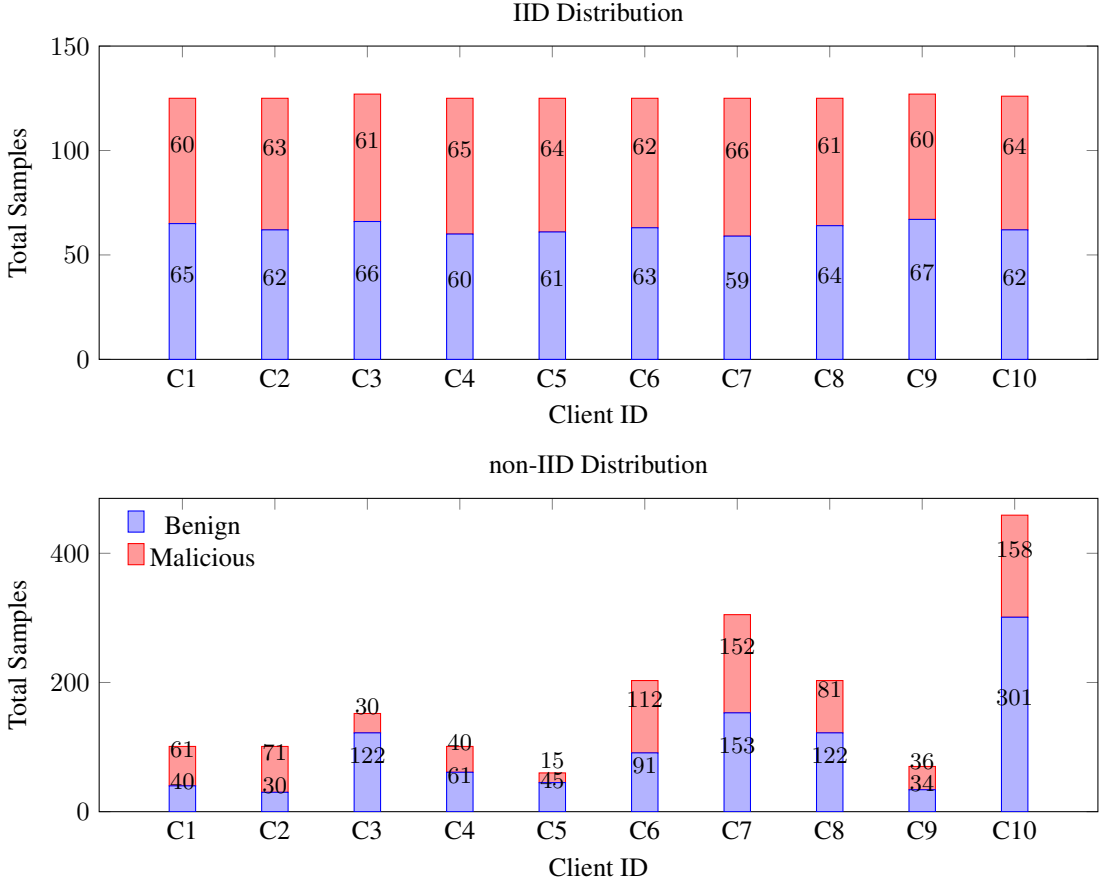
As shown in Figure ~\ref{fig:data_dist}, the IID setup maintains uniform class distribution across clients, while the non-IID setup introduces significant variation in class proportions. This allows for a more realistic and challenging evaluation of federated model convergence and generalization.

\subsubsection{Aggregation and Training Protocol}
Each client locally trained a shared GRU-based model on TF-IDF-transformed system call sequences.
Clients performs local training on their data and shares updated weights with the central server. The server used the FedAvg aggregation protocol to update the global model. The updated global model was then redistributed to all clients for the next communication round. The general workflow of the \textit{AndroIDS} framework is described in Algorithm~\ref{androids}.

The simulation enviroment was setup on local systems, with each client assigned 2 CPU cores, resmembling resource constraint setup as in mobile devices. Experiments were conducted using the same setup for both data distribution scenarios, and to ensure a fair comparison with the centralized setup, the GRU model architecture was kept identical in the FL case.
This experimental setup allows for a flexible evaluation of FL-based intrusion detection under various settings. It serves as the foundation for further experimentation and analysis presented in the next section.
\begin{table}[ht]
\centering
\caption{Explanation of abbreviations used in Algorithm~\ref{androids}.}
\label{tab:abbreviations}
\begin{tabular}{|c|l|}
\hline
\textbf{Symbol} & \textbf{Description} \\
\hline
$\mathcal{M}_g^0$ & Initial global model \\
$\mathcal{M}_g^R$ & Final global model after $R$ rounds \\
$\mathcal{C}$ & Set of all clients \\
$\mathcal{S}$ & Selected subset of clients for training \\
$\mathcal{T}$ & TF-IDF transformer \\
$W$ & Sliding window size \\
$S$ & Stride length for sliding window \\
$\epsilon$ & Standard deviation of Gaussian noise \\
$R$ & Total number of communication rounds \\
$E$ & Number of local training epochs \\
$\mathcal{W}_r$ & Global model weights at round $r$ \\
$\mathcal{W}_c^r$ & Updated local model weights from client $c$ at round $r$ \\
$n_c$ & Number of local samples at client $c$ \\
\hline
\end{tabular}
\end{table}

\begin{algorithm}[ht]
\caption{AndroIDS: FL based system call drive IDS
\\ \small \textbf{Server and Client Side}}
\begin{algorithmic}[1]
\REQUIRE Initial global model $\mathcal{M}_g^0$, client set $\mathcal{C}$, TF-IDF transformer $\mathcal{T}$, window size $W$, stride $S$, noise factor $\epsilon$, number of rounds $R$
\ENSURE Final global model $\mathcal{M}_g^R$
\FOR{round $r = 1$ to $R$}
    \STATE Server selects a subset of clients $\mathcal{S} \subseteq \mathcal{C}$
    \FOR{each client $c \in \mathcal{S}$ in parallel}
        \STATE Send current global model weights $\mathcal{W}_r$ to client $c$
        \STATE Client loads local system call logs $\mathcal{D}_c$
        \STATE Remove timestamps and tokenize syscalls using $\mathcal{T}$
        \STATE Apply sliding window with size $W$ and stride $S$
        \STATE Inject Gaussian noise $\mathcal{N}(0, \epsilon^2)$ into input vectors
        \STATE Initialize GRU model with weights $\mathcal{W}_r$
        \STATE Train model locally on preprocessed data for $E$ epochs
        \STATE Send updated model weights $\mathcal{W}_c^r$ and sample count $n_c$ to server
    \ENDFOR
    \STATE Server aggregates using FedAvg:
    \STATE \quad $\mathcal{W}_{r+1} \leftarrow \left(\sum n_c \cdot \mathcal{W}_c^r \right) \big/ \left(\sum n_c\right)$
    \STATE Update global model: $\mathcal{W}_r \leftarrow \mathcal{W}_{r+1}$
\ENDFOR
\RETURN Final global model $\mathcal{W}_R$
\end{algorithmic}
\label{androids}
\end{algorithm}
\begin{figure}[ht]
    \centering
    \begin{tikzpicture}
        \begin{axis}[
            width=0.48\textwidth,
            height=0.35\textwidth,
            xlabel={Communication Round},
            ylabel={Accuracy (\%)},
            ymin=20, ymax=100,
            xtick={1,2,3,4,5,6},
            ytick={20,30,40,50,60,70,80,90,100},
            grid=both,
            tick label style={font=\small},
            label style={font=\small},
            legend style={font=\small},
            legend pos=south east,
            every axis plot/.append style={thick}
        ]

        \addplot[
            color=blue,
            mark=*,
        ] coordinates {
(1,28.74)
(2,44.55)
(3,56.74)
(4,64.19)
(5,84.32)
(6,96.46)

        };
        \addlegendentry{FL (IID)}

        \addplot[
            color=red,
            mark=square*,
        ] coordinates {
(1,27.29)
(2,42.06)
(3,51.82)
(4,62.54)
(5,81.95)
(6,92.87)
        };
        \addlegendentry{FL (non-IID)}


        \end{axis}
    \end{tikzpicture}
    
    \caption{Global model accuracy for FL schema with 10 clients}
    \label{fig:fl_accuracy_plot}
\end{figure}
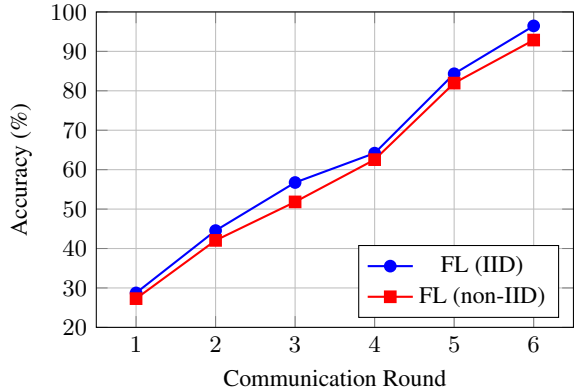
\section{Experimental Results}\label{results}

This section presents the evaluation of our proposed intrusion detection framework using both centralized deep learning and FL setups under IID and non-IID data distributions. Detection performance is assessed in terms of accuracy, precision, recall, and F1-score, and analyze the effect of label skew across clients in FL.
\subsection{Experimental Setup}

All experiments were conducted in a simulated environment on local devices with CPU-only resources. The centralized and federated models share an identical GRU-based architecture. For FL setup, the Flower framework was used with 3 , 6 and 10 clients. Each client was allocated 2 CPU cores and trained locally for 20 epochs per communication round. A total of 6 FL rounds were executed in each experiment. To ensure comparable setups, identical early stopping mechanisms as used in DL were used in FL setup as well.
~Evaluation metrics were computed on a 30\% hold-out validation set.
The results generated for the centralized training set-up is showcased in Table.~\ref{tab1}. 
\begin{table}[ht]
\centering
\caption{Performance Comparison: Centralized vs Federated Learning (IID and non-IID)}
\label{tab1}
\begin{tabular}{lcccc}
\hline
\textbf{Method} & \textbf{Accuracy} & \textbf{Precision} & \textbf{Recall} & \textbf{F1-score} \\
\hline
Centralized DL        & 97.24 & 94.47  & 94.89  & 94  \\
FL (IID)              & 96.46 & 89.43 & 87.67& 89 \\
FL (non-IID)          & 92.87 & 86.43 & 85.13 & 86 \\
\hline
\end{tabular}
\end{table}
\begin{figure*}[ht]
    \centering
    \begin{subfigure}{0.23\textwidth}
        \centering
        \includegraphics[height=4.25cm,width=\linewidth]{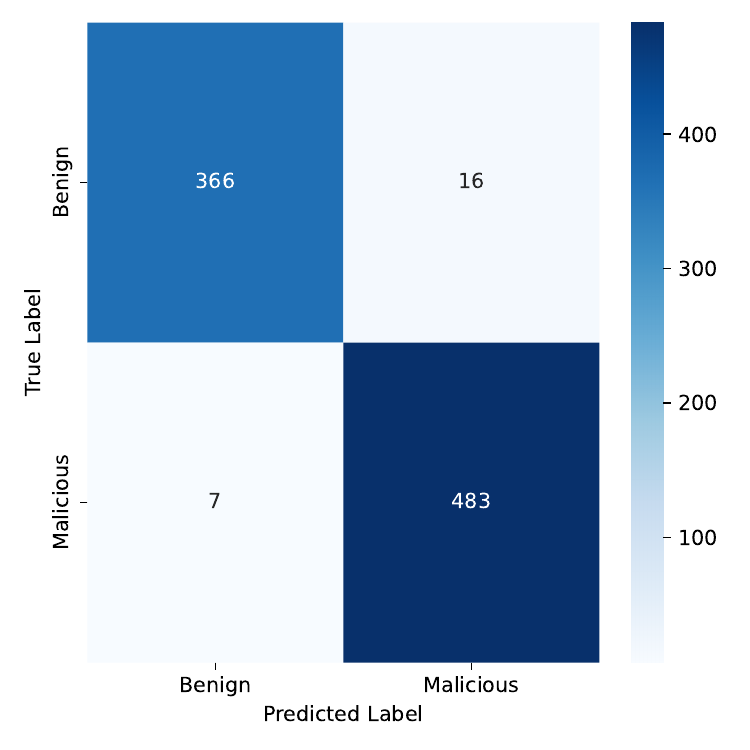}
        \caption{DL model with complete data}
        \label{fig:cm_blnoniid}  
    \end{subfigure}
    \hfill
    \begin{subfigure}{0.23\textwidth}
        \centering
        \includegraphics[height=4.25cm,width=\linewidth]{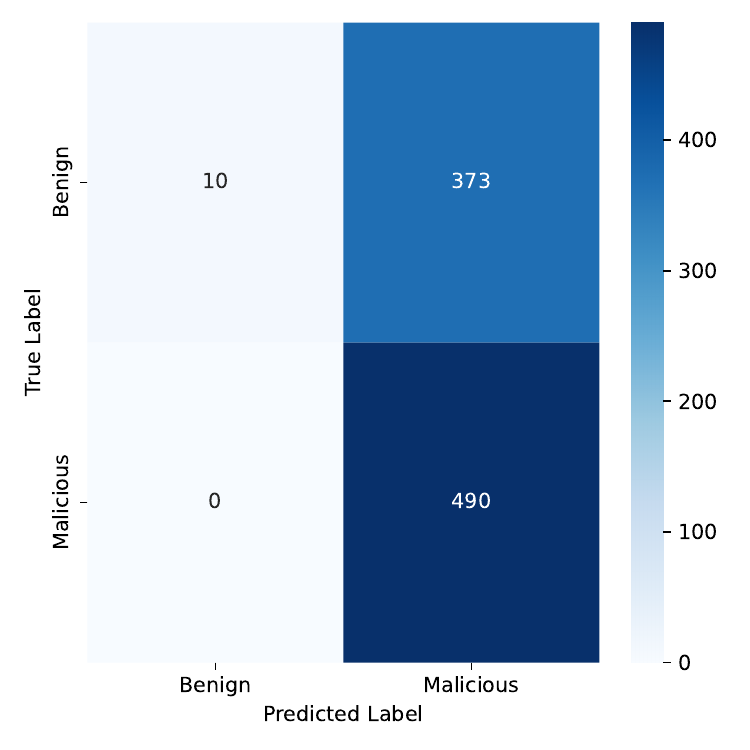}
        \caption{Initial FL global model}
        \label{fig:cm_fliid}
    \end{subfigure}
    \hfill
    \begin{subfigure}{0.23\textwidth}
        \centering
        \includegraphics[height=4.25cm,width=\linewidth]{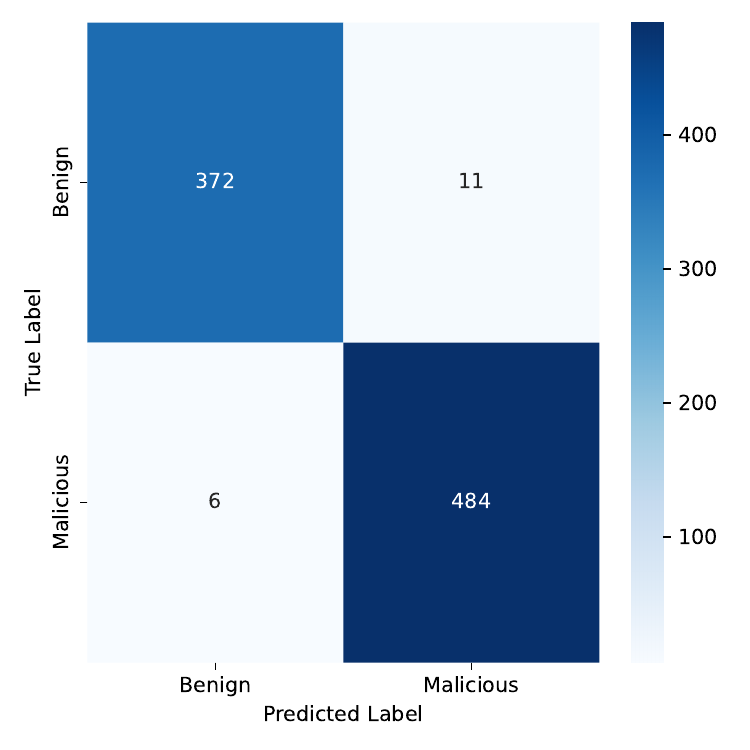}
        \caption{FL model under IID setting}
        \label{fig:cm_bliid}
    \end{subfigure}
    \hfill
    \begin{subfigure}{0.23\textwidth}
        \centering
        \includegraphics[height=4.25cm,width=\linewidth]{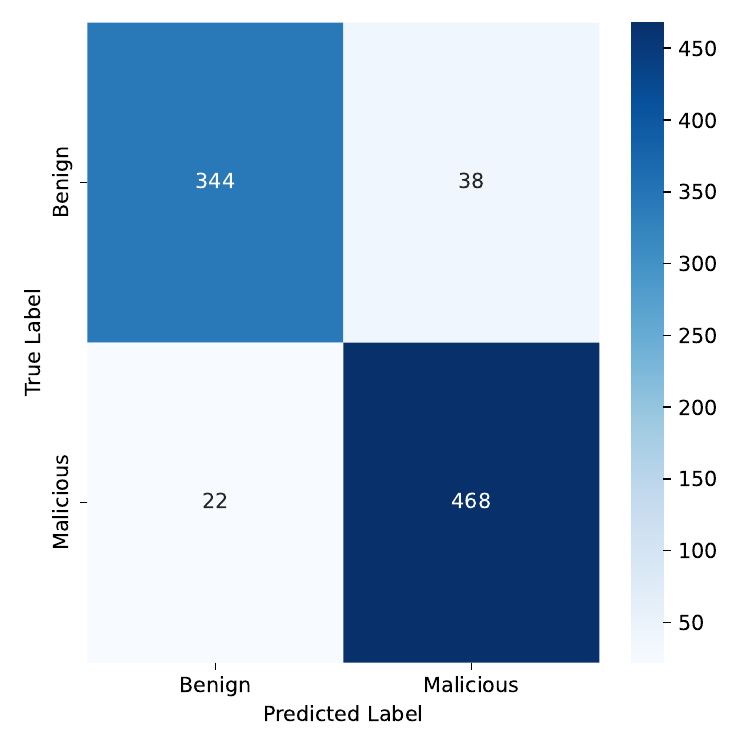}
        \caption{FL model under non-IID setting}
        \label{fig:cm_flnoniid}
        
    \end{subfigure}
    
    \caption{Comparison of confusion matrices for DL model and FL models under IID and non-IID distributions.}
    \label{fig:all_cm}
\end{figure*}
As stated before, FL training was conducted across multiple stages, initially under IID and then under non-IID distribution, with 3, 6, and 10 clients each. The number of clients could not be increased further due to limitations in dataset size, as additional clients would result in smaller partitions, leading to extreme overfitting and model bias.~After six communication rounds, the FL model achieved performance comparable to the centralized baseline (based on early stopping) and the observations are as depicted in Figure~\ref{fig1}.
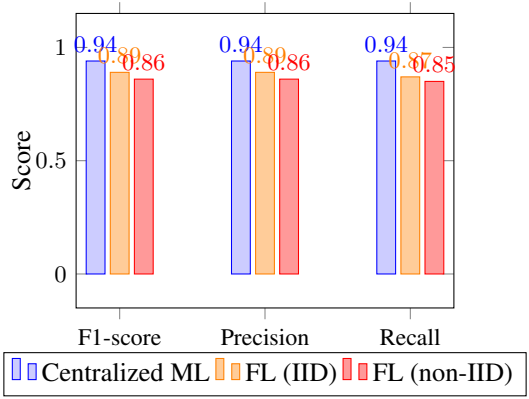
\begin{figure}[ht]
  \centering
  \begin{tikzpicture}
\begin{axis}[
    ybar,
    bar width=.25cm,
    width=0.4\textwidth,
    height=5.5cm,
    enlargelimits=0.15,
    ylabel={Score},
    symbolic x coords={F1-score, Precision, Recall},
    xtick=data,
    ymin=0.0, ymax=1.0,
    legend style={at={(0.5,-0.15)},
      anchor=north,legend columns=-1},
    nodes near coords,
    nodes near coords align={vertical},
    every node near coord/.append style={font=\small},
    ylabel style={yshift=-5pt},
    xlabel style={yshift=10pt},
    tick label style={font=\small},
]
\addplot+[style={blue,fill=blue!20,mark=none}] 
    coordinates {(F1-score,0.94) (Precision,0.94) (Recall,0.94)};
\addplot+[style={orange,fill=orange!40,mark=none}]
    coordinates {(F1-score,0.89) (Precision,0.89) (Recall,0.87)};
\addplot+[style={red,fill=red!40,mark=none}]
    coordinates {(F1-score,0.86) (Precision,0.86) (Recall,0.85)};
\legend{Centralized ML, FL (IID), FL (non-IID)}

\end{axis}
\end{tikzpicture}
  \caption{F1-score, Precision, and Recall comparison across training paradigms.}
  \label{fig1}
\end{figure}

\subsection{Federated Learning under non-IID Distribution}

To evaluate the robustness of our model under real-world conditions, we simulated a non-IID setting where each client received data dominated by a single label class. The observations made in non IID setting is depicted in Figure~\ref{fig:cm_flnoniid}.. Although the non-IID setting introduced moderate degradation in F1-score and recall, the model remained stable over communication rounds. Table~\ref{tabfull} and Figure~\ref{fig:fl_accuracy_plot} presents the detailed performance metrics for both IID and non-IID scenarios and round-wise accuracy trends respectively.

\subsection{Comparative Analysis}

Tables~\ref{tab1} and~\ref{tabfull}  summarizes the performance of all three training paradigms. While the centralized model consistently delivered the highest accuracy, FL setups demonstrated competitive performance while ensuring data privacy and decentralization. Specifically, the non-IID setting led to a moderate drop in recall and F1-score due to class imbalance across clients. From Table~\ref{tabfull}, it can be observed that the disparity in round-wise performance pattern across IID and non-IID settings becomes more evident as the number of clients increases. For systems with a lower number of clients, the divergence among client data distribution tends to be minimal. However, as the client number increases, actual statistical divergence becomes more evident in data distribution, thus justifying the trend.  Figure~\ref{fig1} illustrates the precision, recall, and F1-score comparison across the three configurations.
Although formal experimental comparisons require adequate state-of-the-art (SOTA) baselines, most of the existing works are built around specific public datasets that differ significantly from ours. Since our study focuses on a federated IDS using raw android system call data, limited literature can be identified aligning directly with this context. Moreover, the choice of deep learning architecture critically impacts performance, and applying SOTA frameworks to our dataset would result in unfair comparisons, as those models are optimized for centralized settings or structurally different data. 

\begin{table}[ht]
\centering
\caption{Performance comparison on non-IID and IID scenarios.}
\label{tabfull}
\begin{tabular}{|c|c|*{4}{c}|*{4}{c}|}
\hline
\rotatebox{90}{\textbf{Clients}} & \rotatebox{90}{\textbf{Rounds}} & 
\multicolumn{4}{c|}{\textbf{non-IID}} & 
\multicolumn{4}{c|}{\textbf{IID}} \\
\cline{3-10}
& & 
\textbf{A} & \textbf{P} & \textbf{R} & \textbf{F1} & 
\textbf{A} & \textbf{P} & \textbf{R} & \textbf{F1} \\
\hline
\multirow{3}{*}{3} 
& 2  & 0.44 & 0.41 & 0.40 & 0.40 & 0.48 & 0.42 & 0.40 & 0.42 \\
& 4  & 0.67 & 0.55 & 0.59 & 0.56 & 0.75 & 0.66 & 0.65 & 0.66 \\
& 6  & 0.95 & 0.92 & 0.93 & 0.92 & 0.97 & 0.94 & 0.94 & 0.94 \\
\hline
\multirow{3}{*}{6} 
& 2  & 0.42 & 0.38 & 0.38 & 0.38 & 0.44 & 0.40 & 0.41 & 0.40 \\
& 4  & 0.62 & 0.57 & 0.54 & 0.56 & 0.66 & 0.60 & 0.61 & 0.61 \\
& 6  & 0.94 & 0.90 & 0.88 & 0.88 & 0.97 & 0.93 & 0.94 & 0.94 \\
\hline
\multirow{3}{*}{10} 
& 2  & 0.42 & 0.35 & 0.37 & 0.34 & 0.43 & 0.46 & 0.45 & 0.45 \\
& 4  & 0.62 & 0.63 & 0.59 & 0.60 & 0.64 & 0.63 & 0.67 & 0.64 \\
& 6  & 0.92 & 0.84 & 0.85 & 0.86 & 0.96 & 0.89 & 0.87 & 0.89 \\
\hline
\multicolumn{10}{|c|} {A $\rightarrow$ Accuracy, P $\rightarrow$ Precision, R $\rightarrow$ Recall, F1 $\rightarrow$ F1-score.} \\
\hline
\end{tabular}
\end{table}

The model effectively distinguishes between benign and malicious system call sequences as testified by the confusion matrices for IID and non IID in figures~\ref{fig:cm_bliid} and~\ref{fig:cm_flnoniid} respectively.
While prior works such as ensemble classifiers or graph-enhanced CNNs report high accuracy, they often rely on centralized feature extraction and heavyweight architectures. In contrast, our approach achieves competitive F1-scores of up to \textbf{0.89 (IID FL) and 0.85 (non-IID FL)} using a lightweight GRU-based model with no raw data sharing, making it well-suited for deployment in mobile IoT environments. Thus, the proposed work is positioned as a domain-specific, privacy-aware framework tailored for realistic, resource-constrained federated learning settings.
\section{Discussions}\label{discussions}

Even though non-IID client distribution was employed, the experimental results reveal that the performance of the FL model remains comparable to the IID scenario. This observation is caused by several inherent features of the training framework. First, even though the local client datasets are non-IID, the overall global dataset does not showcase extensive skew across classes. Furthermore, malicious samples outnumber benign samples, thus giving the model sufficient data ensuring that the aggregated gradients efficiently captures representative learning pattern. Another critical factor is the initialization strategy for client selection. It is initialized such that all clients participate synchronously in every training round, which helps smooth out individual client biases and stabilizes global convergence. These two factors contribute highly to the performance of the model. It is important to note that the initial global model at round zero, initialized with random or default weights, performs poorly on the validation data~\ref{fig:cm_fliid}, confirming that no pretraining advantage was present. The convergence, therefore, is driven by the federated training process alone. Further, employing 10 clients ensures standard diversity while maintaining a manageable level of statistical variance. Additionally, the application of Gaussian noise during input preprocessing acts as a regularization mechanism, promoting generalization and reducing overfitting to local distributions. Together, these aspects contribute to narrowing the expected performance gap between IID and non-IID scenarios, resulting in comparable accuracy in the final global model.

 A major limitation of the current \textit{AndroIDS} framework is the lack of comparative validation against SOTA intrusion detection mechanisms. As stated, the framework is positioned as a domain-specific detection model tailored for proprietary android system call data, which justifies the absence of direct comparisons. To address this limitation, we plan to extend our work by integrating the proposed IDS into UAV-based systems, where android-powered devices are increasingly deployed. Furthermore, to overcome challenges related to data integrity and computational constraints in lightweight UAV environments, we aim to enhance the FL schema by exploring more robust hybrid distributed learning alternatives. These future works will be validated using multiple benchmark datasets and real-world noisy environments, thereby addressing the shortcomings identified in the current version.
\section{Conclusion}\label{concluion}

This article presents \textit{AndroIDS}, a FL framework for system call-based intrusion detection in android systems.
The model effectively transforms raw system call sequences into meaningful features via TF-IDF vectorization and employs GRU architecture to capture temporal patterns within the input windows. Experiments conducted across both IID and non-IID client distributions reveal that the proposed framework achieves robust and consistent performance. Under non-IID settings, the model achieved an accuracy of 92.87\%, compared to 96.46\% in the IID scenario, indicating a performance gap of 4\%. Similarly, the F1 score remained comparable, with 86\% under non-IID and 89\% under IID conditions, demonstrating the framework's robustness to data heterogeneity.

The initial global model initialized with random or default weights, performed poorly on the validation data, confirming that no pretraining advantage was present. The convergence, therefore, is driven by the training flow of \textit{AndroIDS} framework alone. The global dataset remains relatively class-balanced, even though local client datasets follow a non-IID distribution. Synchronous participation of all clients in every round helps to stabilize convergence by smoothing out individual client biases. Additionally, the application of Gaussian noise during input preprocessing serves as a regularization mechanism, enhancing generalization and mitigating overfitting. Collectively, these features result in the narrow performance gap between IID and non-IID settings, resulting in comparable accuracy in the final global model.

\bibliographystyle{unsrt}
\bibliography{references}  






\end{document}